\documentclass{article}

    \PassOptionsToPackage{numbers, compress}{natbib}

\usepackage{natbib}
\usepackage{graphicx}
\graphicspath{ {./} }

    \usepackage[final]{neurips_2021}

\usepackage[utf8]{inputenc} 
\usepackage[T1]{fontenc}    
\usepackage{hyperref}       
\usepackage{url}            
\usepackage{booktabs}       
\usepackage{amsfonts}       
\usepackage{nicefrac}       
\usepackage{microtype}      
\usepackage{xcolor}         

\title{Telling Creative Stories Using Generative Visual Aids}

\author{
  Safinah Ali\thanks{Work was conducted while author was at Facebook AI Research.} \\
  MIT Media Lab\\
  Cambridge, MA, USA\\
  \texttt{safinah@mit.edu} \\

   \And
   Devi Parikh \\
Facebook AI Research; Georgia Institute of Technology\\
    Menlo Park, CA, USA\\
   \texttt{parikh@gatech.edu} \\
 
}
\begin{document}
\maketitle
\begin{abstract}
  Can visual artworks created using generative visual algorithms inspire human creativity in storytelling? We asked writers to write creative stories from a starting prompt, and provided them with visuals created by generative AI models from the same prompt. Compared to a control group, writers who used the visuals as story writing aid wrote significantly more creative, original, complete and visualizable stories, and found the task more fun. Of the generative algorithms used (BigGAN, VQGAN, DALL-E, CLIPDraw), VQGAN was the most preferred. The control group that did not view the visuals did significantly better in integrating the starting prompts. Findings indicate that cross modality inputs by AI can benefit divergent aspects of creativity in human-AI co-creation, but hinders convergent thinking.
\end{abstract}



\section{Introduction}

Generative AI algorithms have found applications in visual art, stylizing images and videos. Combined with Contrastive Language–Image Pre-training (or CLIP) by OpenAI~\cite{radford2021learning}, that learns visual concepts from natural language supervision, it has become possible to leverage a combination of generative approaches and CLIP to create images from language prompts. Repeated generation until the image sufficiently represents the prompt as scored by CLIP eventually creates a visual that is semantically relevant to the prompt. This approach of combining CLIP with GANs such as StyleGAN, BigGAN, VQGAN, etc. to create visuals from textual prompts has led to a recent trend in creative technologists and artists sharing surreal AI-generated art from language prompts~\cite{snell, murdock}.

We study whether artwork created using generative algorithms in one medium can inspire creativity in another. Artists find creative inspiration from external stimulants, even across mediums~\cite{duncum2004visual}. Musician Debussy’s La Mer was inspired from The Great Wave~\cite{zinn}. Graphics have been used to inspire reading, writing and thinking~\cite{claggett1992drawing, childers1998articulating}. Teachers have found that using manually curated pictures helped students’ narrative building ability and interest in writing ~\cite{ali2014using, asrifan2015use}. Here, we study whether art created using GANs and other visual generative procedures help inspire creative storytelling. As opposed to manually curating artwork, generative algorithms automate the process of generating images contextual to a language prompt. Further, it is difficult to manually curate (e.g. using search engines) visuals for unrealistic fictional prompts, such as “an octopus playing a Ukulele”, but they can be generated using generative approaches. Generative art also contains surrealistic elements, which could introduce unexpected elements for the writer, which could influence the stories.

\section{Methods}
We designed \textbf{Visual Stories}, a web interface where writers write a creative story from a starting prompt. They are aided with visuals generated using the same prompt for creative inspiration, using: (1) CLIP + BigGAN~\cite{brock2018large, bigsleep}. (2)  CLIP + VQGAN (Vector Quantized Generative Adversarial) Networks~\cite{esser2021taming}. (3) CLIPDraw (stroke-based drawings)~\cite{frans2021clipdraw}. (4) CLIP+DALL-E (Decoder of the discrete VAE used for DALL-E) ~\cite{ramesh2021zero}.

\begin{figure}
  \centering
  \includegraphics[width=0.87\textwidth]{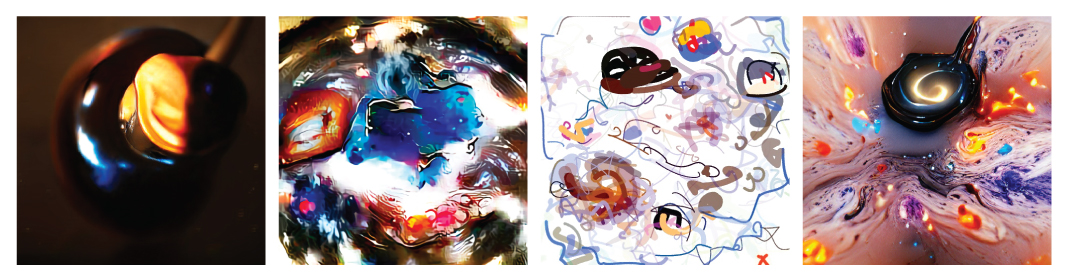}
  \caption{Visuals used for the starting prompt “A galaxy in a melting pot.” generated using BigGAN, DALL-E decoder, CLIPDraw and VQGAN.}
  \label{interface}
\end{figure}

To understand whether the generative visuals benefited writer's creativity, we ran a human experiment, where adult participants who are fluent in English were recruited from Amazon Mechanical Turk (AMT) to write creative stories from fixed starting prompts. Participants were randomly divided into two groups: one that could toggle between four visuals generated from the starting prompt they were provided (V+), and one that saw no visuals (V-). After the task, participants reflected on their writing experience, creativity of the stories, and usefulness of the visuals (~\ref{questionnaire}). The story writing web interface consisted of instructions (\ref{instructions}), a starting prompt for the stories, a text box, a button to view and toggle through visuals (Figure~\ref{interface}). External judges blind to the conditions rated the stories using metrics from the “convergent integrative thinking” segment for verbal creativity of the Evaluation of Potential Creativity (EPoC) battery~\cite{barbot2016generality} (~\ref{scoring}).

\section{Findings}
362 stories were analyzed. Participants in the V+ condition, scored the task as more fun (4.126 ± 0.908) than participants in the V- condition (4.307 ± 0.808) (p = 0.0453). They also wrote longer stories (828 characters as compared to 796 characters in V-). 65\% writers found the visuals inspiring. Images generated using VQGAN were the most frequently used images, and VQGAN, BigGAN and CLIPDraw generated images were used significantly more than DALL-E (p = 0.0048). The visuals inspired “characters”, “setting” and “details” of the stories the most.     

A chi-square test of goodness-of-fit test was performed to determine whether the stories written by participants in the two study conditions were equally rated on the specified metrics of creativity by external raters. Rejecting the null hypothesis, preference for the two categories was not equally distributed in the population, and rated higher for the intervention group on the metrics of “\textit{Creativity}”, “\textit{Originality}”, “\textit{Level of completion}”, “\textit{Visualizability}”, and “\textit{Plot}” (\textit{p<0.05}). Preference for “\textit{Integration of prompt}” was rated significantly higher for the control group (\textit{p<0.05}). 

We found that the generated images influenced people’s creative process and outcome. They found the process more fun and the graphics helpful. External evaluation indicated that it benefited writers’ divergent-exploratory thinking, and the stories were more creative and original. However, it hindered convergent-integrative thinking - the participants did not integrate the starting prompt as well as the control group. This could be caused by participants being distracted by the images, or follow the arc of the images and use the visuals as the prompt, steering away from the starting textual prompt. This could also indicate that perhaps the visual images were not sufficiently relevant to the starting prompt. Stories from the V+ condition were preferred by raters to be converted to an illustration or animation, implying that the stories helped the reader visualize it, influencing not just the writing process, but also the reading process. 

Generative approaches coupled with CLIP allow the creation of high quality visuals that are able to visualize unrealistic scenarios. Different approaches also bring unique aesthetic styles. While these are currently shared as digital artwork, in this work we explore how these can also be used as effective creative writing aids. Future work could explore how they could aid creativity in other media, such as music production. Not requiring manual curation makes them effective automated aids for creative writing. Instead of static images pre-generated from text, future work could use parts of the written story to dynamically update the aesthetics of the generation. 


\section{Ethical Implications}
In this work, we demonstrate how visual art generated by AI was a helpful writing aid. However, we also found that the writers veered away from the starting textual prompt, and towards the visuals. A cross medium AI collaborator helped with divergent thinking, but hindered convergent integrative thinking, and while designing creative collaborative AI, we must reflect on the cost it may potentially have in hindering the task at hand. There is also the risk of creativity aids becoming dependencies. For instance, one writer reflected that before the visuals, they were not sure what to write. Relying on these visual aids may hinder their verbal creativity when the visuals are missing. 

\bibliography{sample}

\section{Appendix}

\subsection{Story Writing Instructions} \label{instructions}

\begin{itemize}
    \item In this task, you will be writing a creative story.
    \item Use the given starting sentence and continue it to create your original creative story.
    \item Write a minimum of 500 characters (about 100 words) and maximum of 1000 characters to complete the task.
    \item You can choose to view some visual inspiration to help with the story using the yellow button on the top.
    \item Most creative stories will be awarded a bonus.
    \item Please do not paste text from the internet and try to create your own story!
    \item Please do not complete this HIT multiple times. It will only be accepted once.
\end{itemize}

\subsection{Story scoring} \label{scoring}

We adopted story rating metrics from the “convergent integrative thinking” segment for verbal creativity of the Evaluation of Potential Creativity (EPoC) battery~\cite{barbot2016generality} as well as the Consensual Assessment Technique (CAT)~\cite{hennessey1999consensual}. More details of EPoC, CAT and the questions asked can be found in Appendix 1. 

The EPoC test was chosen since it involves a verbal creativity assessment task similar to the task we used, where subjects are asked to create a story from a fixed beginning. While the EPoC is developed for children, it has been used with adults. The CAT is a powerful tool used by creativity researchers in which judges are asked to rate the creativity of creative products such as stories, collages, poems, and other artifacts. Creativity is rated through comparisons with other artifacts in the pool, and not as absolute scores for each individual story. The CAT has been used in many contexts to compare the creativity of two groups. Raters were asked to compare the stories on the following metrics: 

\begin{itemize}
    \item Which of these stories is more creative (or is more novel or original)?
    \item Please explain your answer in 1-2 sentences
    \item The writer of which story is more imaginative?
    \item Please explain your answer in 1-2 sentences
    \item Which story better integrates the starting prompt provided?
    \item Which story has a more elaborate setting (or has the setting explained in detail)?
    \item Which story has more elaborate characters (or has characters explained in detail)?
    \item Which story has a better narrative composition or storyline?
    \item Which story was more interesting for you to read?
    \item Which story do you think has an overall higher quality?
    \item Which story seems more complete?
    \item Hypothetically, if you had to make an illustration, visual image or animation from one of these stories, which one would you rather pick?
\end{itemize}

\subsection{Post-test questionnaire} \label{questionnaire}
Participants were asked the following questions after writing their stories: 
\begin{itemize}
\item Did the visual images inspire you for creative story writing?
\item On a scale of 1-5, how fun was the writing process?
\item Would you want to share the story you wrote with your friends?\textit{ [Yes, No, Maybe]}
\item Do you think the story you wrote was creative? \textit{[Yes, it was very creative; Yes, but I can do better; It was not great, but not bad either; It was not that creative; No, it was not creative at all].}
\item In 1-2 sentences, describe how the visuals inspire you for writing.
\item In 1-2 sentences, describe what you wish the website could provide that could better help you with writing.
\item Which parts of the story did the visuals help you with? \textit{[Characters, setting, details, plot, conflict, resolution, tone, none]}
\end{itemize}

\subsection{Study findings} \label{graphs}

\begin{figure}[h]
  \centering
  \includegraphics[width=\textwidth]{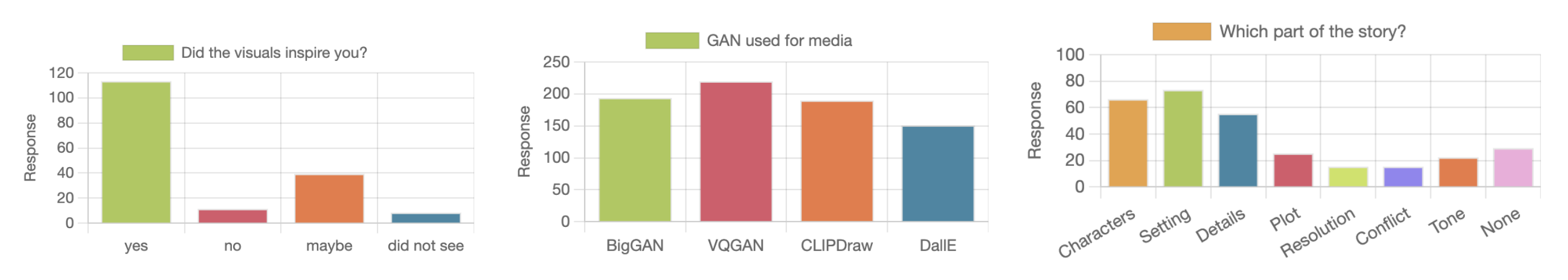}
  \caption{65\% writers found the visuals inspiring. Images generated using VQGAN were the most used images, and VQGAN, BigGAN and CLIPDraw generated images were used significantly more than DALL-E generated images (X2 (3, N = 751) = 12.947 , p = 0.0048). Of the features of stories that the visuals inspired, “characters”, “setting” and “details” were the most common. }
  \label{graphs}
\end{figure}

\subsection{Sample stories}

\subsubsection{With the visual prompt (Figure~\ref{interface})}
"\textbf{A galaxy in a melting pot} was the first impression Betsy had of the abstract painting. And yet, as she studied it more closely, the swirling blue black cyclone and bright star-like spots began to seem almost familiar to her. It was as if bits here and there had attached to her psyche as snippets of memories she had all but forgotten. She continued to observe, and as she did, the painting seemed to awaken from its inert state and began to move. It started slowly but then picked up speed. Betsy was mesmerized. She became frightened but was unable to pull away. More and more invisible tendrils clung to her. She gasped. A tear trickled out of her eye. And suddenly the painting pulled her into itself."

\subsubsection{Without the visual prompt}
"\textbf{A galaxy in a melting pot} was what the Captain called it. And we were tasked with the mission of saving those on surrounding planets. It was a gargantuan task for any society and we had only 1000 ships at our disposal, which would convey only half the surrounding planets population before the likely destruction caused by the supernova. Some planets instituted a lottery system, with families with children at the top of the list. Others were less organized as there were crowds vying to get on the ships and crowd control had to be instituted. But there was one planet that was unique. They refused to let anyone go unless everyone could go. Hearing this, the federation of planets decided to put them at the head of the line and insure that everyone could go on their planet. To accomplish this, they had do de-prioritize the planet where everyone was competing against each other to board the ships. In the end, that planet would be the last one served."

\end{document}